%% file: globecom_main.tex
\documentclass[conference]{IEEEtran}
\usepackage[printonlyused]{acronym}
\usepackage{adjustbox}
\usepackage[style=ieee,backend=bibtex]{biblatex}
\bibliography{biblio}
\include{macros}
\usepackage{array}

\usepackage[font=small]{caption}
\usepackage{subfigure}
\usepackage{bm}
\usepackage{amsmath}
\usepackage{mathtools}
\usepackage{breqn}

\usepackage{textcomp}

\begin{document}

\title{Short-Term Trajectory Prediction for Full-Immersive Multiuser Virtual Reality with Redirected Walking}
\author{
\IEEEauthorblockN{Filip Lemic\IEEEauthorrefmark{1}\IEEEauthorrefmark{2}, Jakob Struye\IEEEauthorrefmark{2}, Jeroen Famaey\IEEEauthorrefmark{2}}
\IEEEauthorblockA{\IEEEauthorrefmark{1}NaNoNetworking Center in Catalunya (N3Cat), Polytechnic University of Catalonia, Spain}
\IEEEauthorblockA{\IEEEauthorrefmark{2}Internet Technology and Data Science Lab (IDLab), Universiteit Antwerpen - imec, Belgium \\
Email: \{filip.lemic, jakob.struye, jeroen.famaey\}@uantwerpen.be}
\vspace{-6mm}
}

\maketitle

\begin{abstract}

Full-immersive multiuser \ac{VR} envisions supporting unconstrained mobility of the users in the virtual worlds, while at the same time constraining their physical movements inside VR setups through redirected walking. 
For enabling delivery of  high data rate video content in real-time, the supporting wireless networks will leverage highly directional communication links that will “track” the users for maintaining the \ac{LoS} connectivity. 
\acp{RNN} and in particular \ac{LSTM} networks have historically presented themselves as a suitable candidate for near-term movement trajectory prediction for natural human mobility, and have also recently been shown as applicable in predicting VR users' mobility under the constraints of redirected walking.  
In this work, we extend these initial findings by showing that \ac{GRU} networks, another candidate from the RNN family, generally outperform the traditionally utilized LSTMs.
Second, we show that context from a virtual world can enhance the accuracy of the prediction if used as an additional input feature in comparison to the more traditional utilization of solely the historical physical movements of the VR users.
Finally, we show that the prediction system trained on a static number of coexisting VR users be scaled to a multi-user system without significant accuracy degradation.

\end{abstract}


\input{acronym_def}
\input{introduction}

\input{system}

\input{results}

\input{conclusion}
\section*{Acknowledgments}
Filip Lemic was supported by the EU H2020 Marie Skłodowska-Curie project ”Scalable Localization-enabled In-body Terahertz Nanonetwork” (nr. 893760).
\vspace{-1mm}
\renewcommand{\bibfont}{\footnotesize}
\printbibliography

\end{document}

%% file: acronym_def.tex

\acrodef{IoT}{Internet of Things}
\acrodef{ISM}{Industrial, Scientific and Medical}
\acrodef{LPWAN}{Low-Power Wide Area Network}
\acrodef{REM}{Radio Environmental Map}
\acrodef{SNR}{Signal-to-Noise Ratio}
\acrodef{MT}{Mobile Terminal}
\acrodef{LoS}{Line-of-Sight}
\acrodef{NLoS}{Non-Line-of-Sight}
\acrodef{MCS}{Modulation and Coding Scheme}
\acrodef{GPS}{Global Positioning System}
\acrodef{RSSI}{Received Signal Strength Indication}
\acrodef{MAE}{Mean Absolute Error}
\acrodef{SE}{Squared Error}
\acrodef{SDR}{Short-Range Device}
\acrodef{QoE}{Quality of Experience}
\acrodef{VR}{Virtual Reality}
\acrodef{AP}{Access Point}
\acrodef{HMD}{Head-Mounted Device}
\acrodef{IRS}{Intelligent Reflective Surfaces}
\acrodef{AoA}{Angle of Arrival}
\acrodef{RNN}{Recurrent Neural Network}
\acrodef{LSTM}{Long Short-Term Memory}
\acrodef{GRU}{Gated Recurrent Unit}
\acrodef{APF-RDW}{Artificial Potential Field Redirected Walking}
\acrodef{APF-R}{Artificial Potential Field Reseting}
\acrodef{HD}{High Definition}

%% file: introduction.tex
\section{Introduction}

\acf{VR} continues to revolutionize our digital perceptions and interactions~\cite{munoz2020augmented}. 
It also promises novel applications with utility in healthcare, tourism, education, entertainment, and occupational safety~\cite{jensen2018review,izard2018virtual}, to name a few. 
VR setups and contents are being rapidly improved, with primary efforts targeting the enhancements in the immersiveness of VR experiences. 
Toward this end, the research is mainly focused on enhancing the quality of video content that is being presented to the VR users~\cite{zhang2019wireless}, as well as  on ``cutting the wire'' and enabling truly wireless delivery of the video content~\cite{chen2018virtual,struye2021millimeter}. 
An additional important objective is to enable multiuser experiences, in which the users are able to collaborate in a way that the action of one user in the virtual world is delivered as a virtual content to other, potentially collocated users~\cite{bachmann2019multi}, and consequently affects their virtual experiences. 

In the near future, VR systems will support multiple fully immersed and coexisting VR users with constraint-free mobility in the virtual worlds.
Such setups will be supported through high frequency wireless communication networks primary operating in the millimeter Wave (mmWave) (i.e., 30-300 GHz) frequency band~\cite{elbamby2018toward}. 
For supporting the delivery of high-quality video content to the mobile VR users in real-time, the underlying wireless communication will have to be highly directional on both transmit and receive sides~\cite{liu2021learning}. 
These directional mmWave beams are simultaneously expected to ``track'' the users’ movements for continuously maintaining \acf{LoS} connectivity for maximizing the quality of video delivery.
At the same time, redirected walking is envisioned to be utilized for avoiding collisions between the collocated users, and between the users and the boundaries of the constrained VR environments~\cite{bachmann2019multi}. 
This will enable immersion of the VR users in the virtual worlds by allowing them to roam freely, while simultaneously and ideally imperceivably redirecting them in the physical setups for collision avoidance.

For supporting continuous LoS maintenance with each of the VR users, the directional mmWave beams will have to be transmitted in such a way that they also provide coverage with high-quality wireless signals at near-future locations of the VR users.
This in turn motivates the need for VR users' short-term movement trajectory prediction in full-immersive multiuser VR setups.
Short-term movement trajectory prediction for a natural human walk is a well-established topic in the research community, which yielded \acp{RNN} in general and \ac{LSTM} networks as their particular \ac{RNN} instance suitable for this task (e.g.,~\cite{song2020pedestrian,bartoli2018context}).
However, it is intuitive that neither imperceivable nor perceivable re-steers of the VR users resemble a natural human walk, suggesting the need for evaluating the suitability of \acp{RNN} for predicting VR users' mobility under the constraints stemming from redirecting walking.
This topic has received substantially less attention in the research community.
Nonetheless, an early work from Nescher and Kunz~\cite{nescher2012analysis} has shown that such precision is possible with reasonable accuracy, while~\cite{hou2019head,cho2018path} demonstrated that \acp{RNN}, specifically LSTMs, can be utilized for this purpose as they feature encouraging prediction accuracy.
All three approaches base the prediction of the near-future trajectories on the historical movement trajectories of the VR users. 

In this work, we extend these initial findings along several dimensions.
First, we consider \acp{GRU} in addition to \acp{LSTM} for the near-future movement trajectory prediction. 
We consider \acp{GRU} due to their faster training and execution in comparison to LSTMs, which would conceptually make them a more suitable candidate for such prediction in case they feature comparable prediction accuracies.
Second, we consider information from the virtual world (i.e., the VR users' movement trajectory in the virtual world) as an additional input feature for the two considered approaches, which is in contrast to the existing works from the literature that base the prediction solely on historical movement trajectories in a physical space. 
Finally, we evaluate the effects of a varying number of coexisting VR users (e.g., the users dynamically entering and terminating immersive experiences) on the prediction performance of the considered approaches trained with the data from only a two-user system.

Our results demonstrate that GRU outperforms LSTM not only in terms of the execution times, but also in the prediction accuracy, suggesting their utility in the considered scenario.
Moreover, we show that the information from the virtual experience, in our case the virtual movement trajectory, provides additional useful context and, therefore, improves the accuracy of prediction for both LSTM and GRU-based approaches.
Finally, our results show that a predictor trained on data from two-user systems only can be scaled to a multi-user system without significant accuracy degradation, which has a potential for enabling dynamic immersion in and termination of virtual experiences in multiuser full-immersive VR setups.

%% file: system.tex
\section{System Overview}

\subsection{Considered Scenario}

\begin{figure}[!t]
\centering
\includegraphics[width=\linewidth]{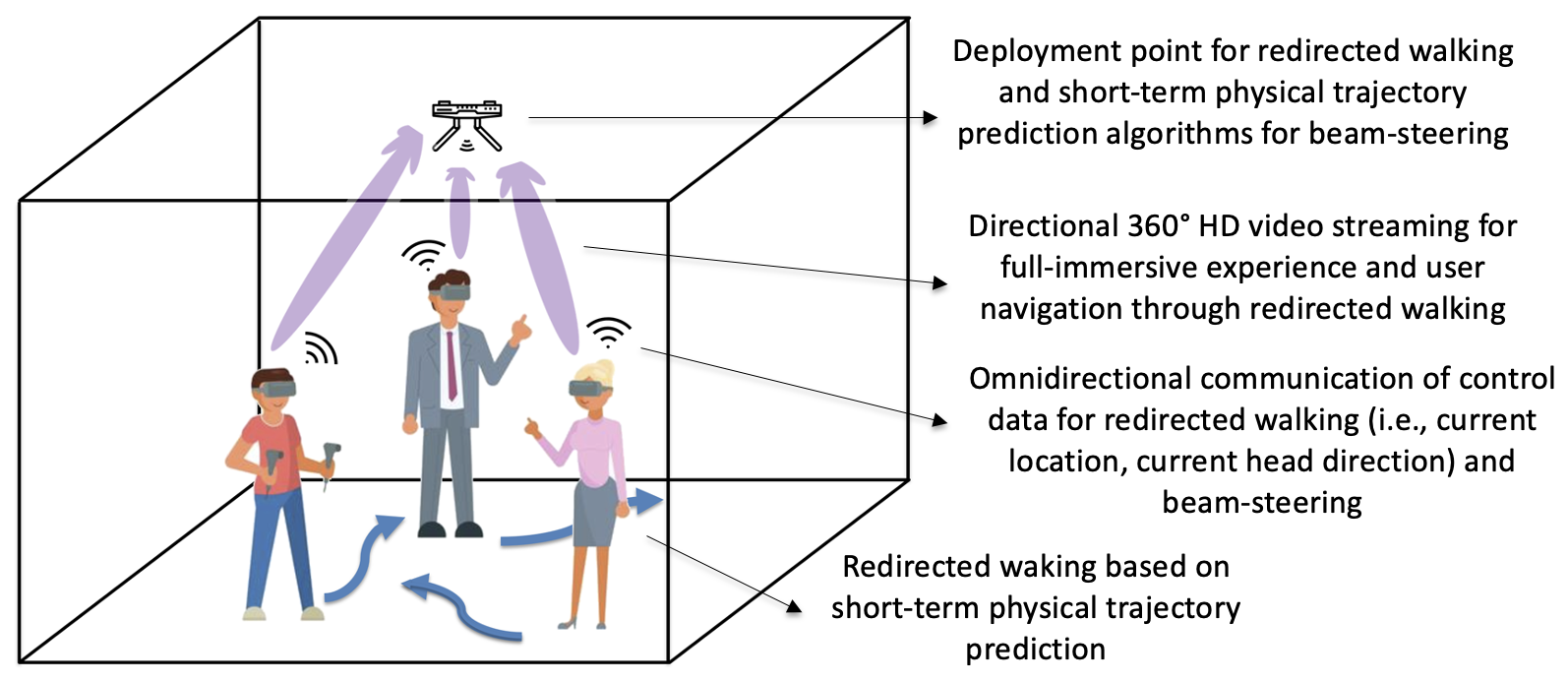}
\caption{Considered scenario}
\label{fig:scenario}
\vspace{-4mm}
\end{figure}

We consider a full-immersive multiuser \ac{VR} scenario as depicted in Figure~\ref{fig:scenario}. 
Specifically, we consider a constrained physical environment in which the full-immersive \ac{VR} setup is deployed.
The environment is constrained in terms of its physical sizes in order to create a boundary inside of which it is safe to engage in virtual experiences.
As such, it is assumed that there are no obstacles inside the deployment environment, which would represent a tripping hazard if not properly accounted for.
In other words, the only factors to be considered as potential collision perils from the perspective of a VR user are the environmental boundaries and other users, as depicted in the figure. 

In the scenario, multiple VR users can be collocated in the same deployment environment. 
The VR content is streamed to the users using highly directional mmWave communication.
Specifically, highly directional beams are transmitted by an \ac{AP} in such a way that they follow the movements of the VR users.
This is done in order to continuously maintain \ac{LoS} connectivity with each of the users, resulting in a maximized link quality and consequently enhancing the VR users' \ac{QoE}~\cite{struye2020towards}.
To do so, we envision the VR users' \acp{HMD} reporting their physical locations to the AP, which is then utilized for supporting both redirected walking and beamsteering.
The assumption of the availability of physical locations of the VR users is well-established in the existing literature (e.g.,~\cite{liu2021learning}) and the contemporary VR headsets such as Oculus Quest 2 and Vive Cosmos already support its generation and provisioning.  
It is also worth emphasizing that these VR headsets are able to support the generation and provisioning of VR users' location information without the support of external devices such as cameras or localization anchors. 

In terms of transmitter-side beamsteering, the AP is envisioned to utilize the current and near future locations of a VR user for forming a beam in a way that it covers both current and near future locations of the user, enabling LoS maintenance during a current time instance, as well as in the near future instances.
Simultaneously, redirected walking is utilized for steering of the VR users in a way that guarantees collision avoidance between the collocated users, as well as between the users and the environmental boundaries.
The aim of redirected walking is to enable immersion of the VR users (i.e., allowing them to roam freely in potentially unconstrained virtual worlds), while simultaneously and (ideally) imperceivably redirecting them in the physical space.

Note that on the receiver side, beamforming is envisioned to adapt in real-time to the VR user's head rotations, relying on the HMD's built-in sensors providing accurate orientation estimates. 
The receiver-side beamforming and beamsteering is considered as out of scope of this work. 
Interested readers can find more details on this aspect of the system in~\cite{struye2021millimeter}, where we present coVRage, a receiver beamforming solution in which, based on past and current head orientations, the HMD predicts how the \ac{AoA} from the \ac{AP} will change in the near future, and covers this AoA trajectory with a dynamically shaped beam.  

Note that we consider as out of scope the potential interruptions of the LoS connectivity between the AP and the users due to the obstructions caused by the other users.  
An intuitive solution to the issue can be based on an AP handover if it is predicted that one user will block the LoS path of another one, again motivating the need for short-term prediction of the VR users' movements.    
Alternative solutions based on \acp{IRS} have also been presented~\cite{liu2021learning}.

\subsection{Short-Term Trajectory Prediction}

\begin{figure*}[!t]
\vspace{-1mm}
\centering
\subfigure[RNN structure]{
\includegraphics[width=0.15\textwidth]{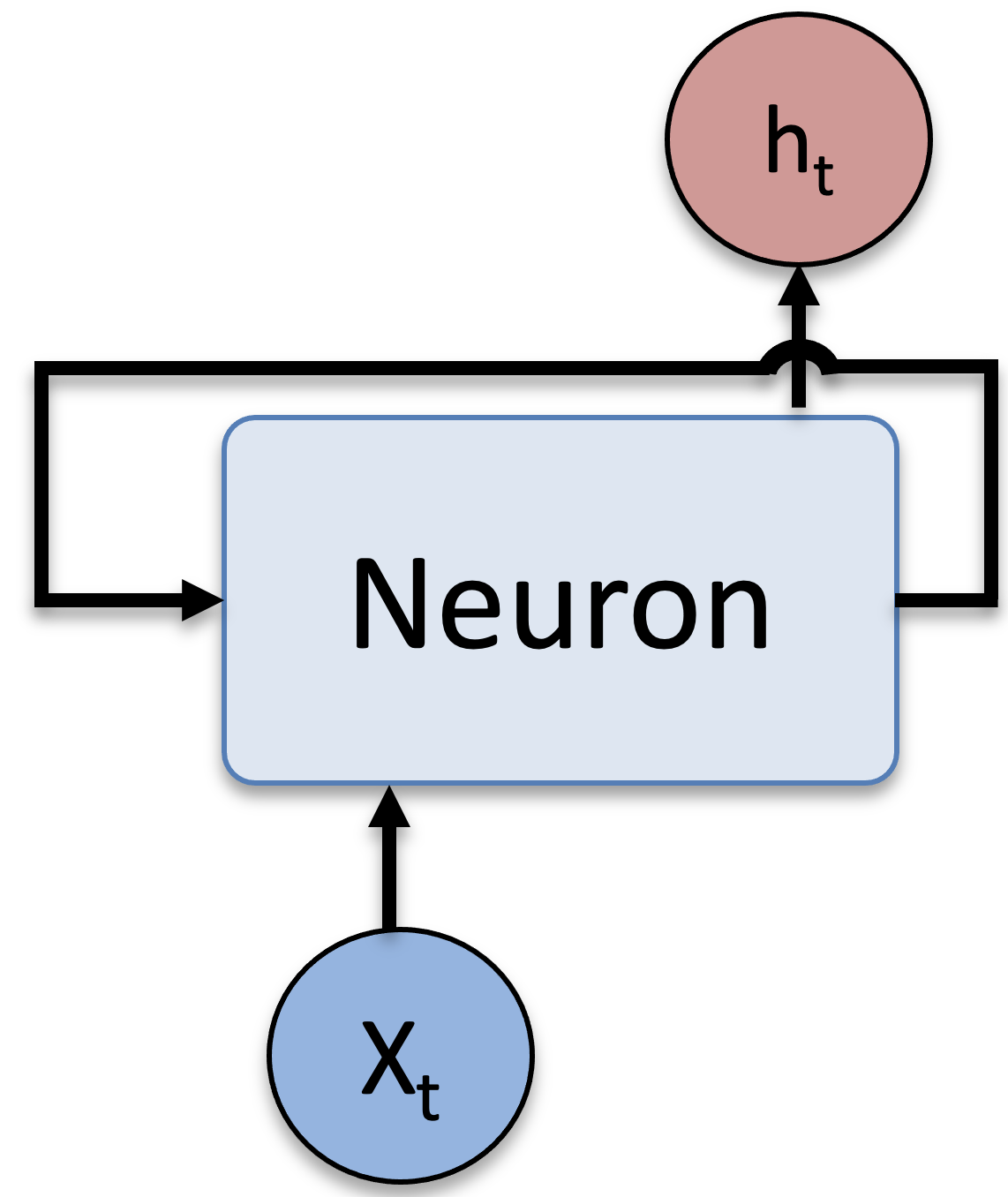}
\label{fig:rnn}}
\subfigure[LSTM neuron]{
\includegraphics[width=0.29\textwidth]{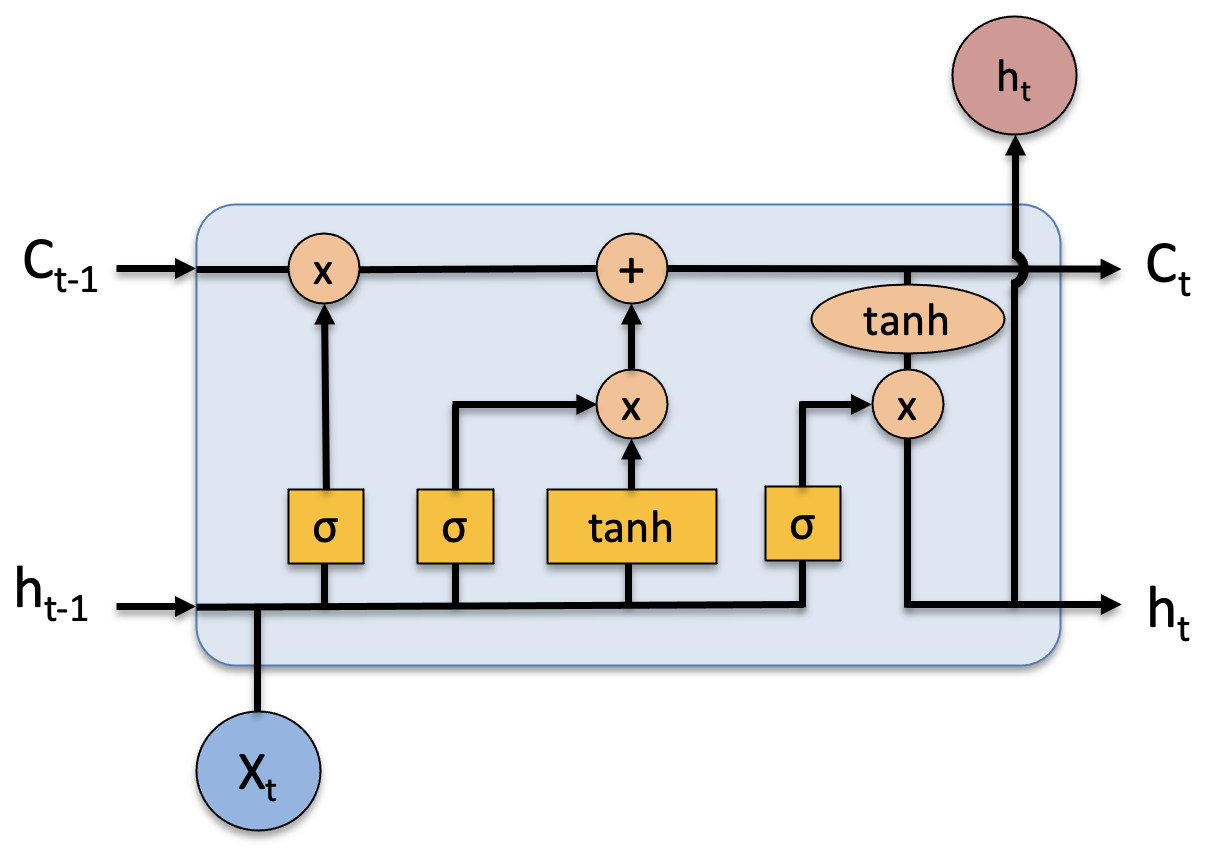}
\label{fig:lstm_neuron}}
\subfigure[GRU neuron]{
\includegraphics[width=0.29\textwidth]{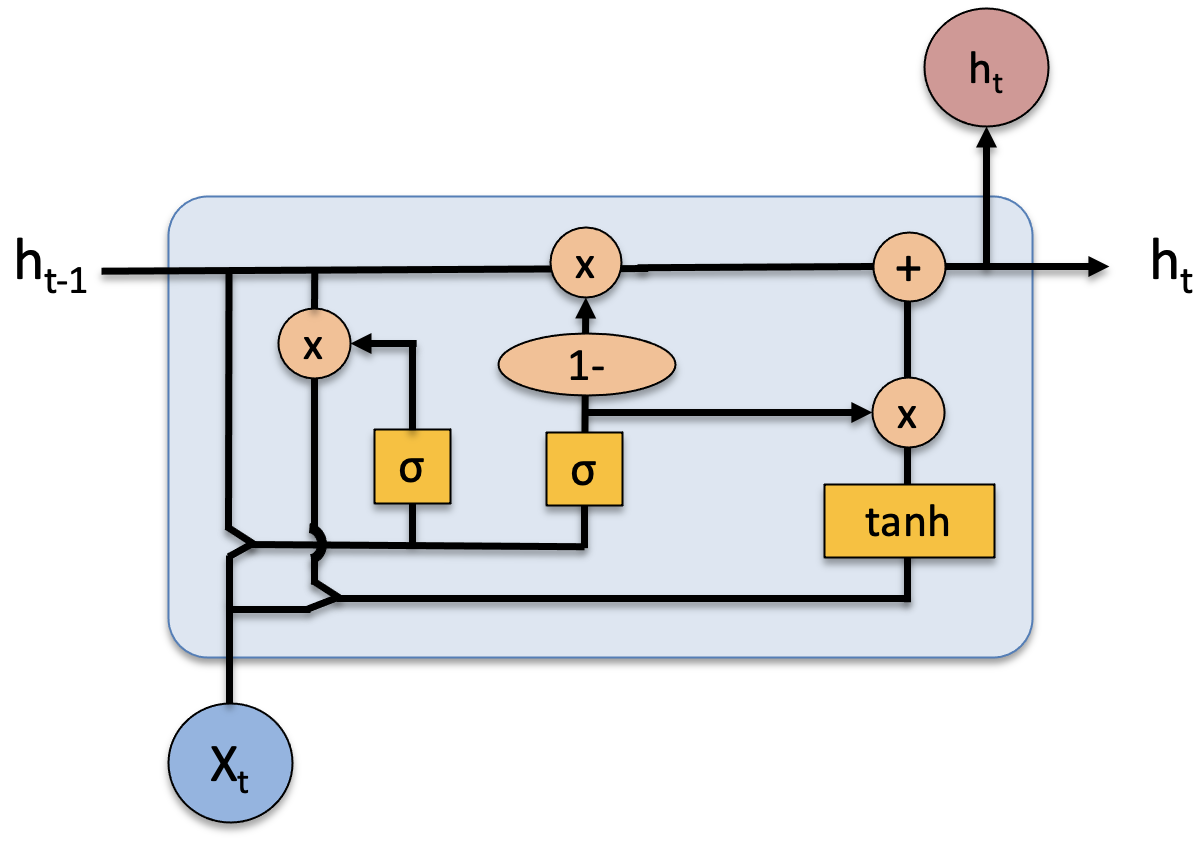}
\label{fig:gru_neuron}}
\subfigure[Legend]{
\includegraphics[width=0.18\textwidth]{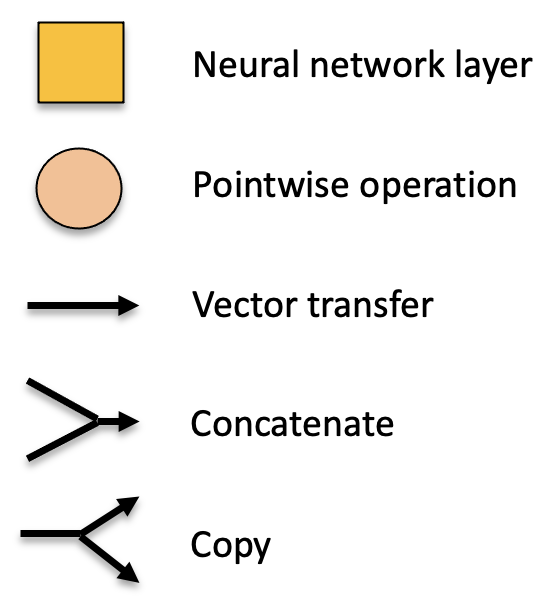}
\label{fig:legend}}
\vspace{-2mm}
\caption{Considered types of recurrent neural networks}
\label{fig:rnns}
\vspace{-5mm}
\end{figure*} 

Based on the above discussion, short-term prediction of the VR users' movement trajectory is envisioned to be utilized for both redirected walking and transmitter-side beam steering, while also being potentially beneficial for LoS obstruction avoidance.
It is intuitive that the accuracy of such prediction will affect the performance of other parts of the envisioned system, suggesting that the optimization of this accuracy will be beneficial from the perspective of the VR's \ac{QoE}.  

As mentioned before, \acp{RNN} are arguably the most suitable technological candidates for predicting the trajectory of human movements, even in redirected walking scenarios. 
An \ac{RNN} is a class of artificial neural networks which contains multiple neurons of the same type, each passing a message to a succeeding neuron, as depicted in Figure~\ref{fig:rnn}.
This allows the RNNs to exhibit temporally dynamic behavior, making them suitable for tasks such as handwriting or speech recognition and time-series prediction~\cite{yu2019review}.
In this work, we consider two of the most promising types of RNNs for such prediction, namely \ac{LSTM} and \ac{GRU}.
Single neurons of both types of networks are depicted in Figure~\ref{fig:rnns}, together with the indications of their main building blocks.

In LSTM, a \emph{sigmoid} layer called the forget gate is utilized for deciding on the retainment of the previous cell state. 
Moreover, a \emph{sigmoid} layer called the input gate jointly with a \emph{tanh} layer are utilized for updating the cell state.
The output of the cell is then based on our cell state filtered through a sigmoid layer for deciding the part of the cell state to be outputted, normalized through a \emph{tanh} layer, as depicted in Figure~\ref{fig:lstm_neuron}.
GRU neuron is comparable with the one from LSTM, with the main differences stemming from the combined forget and input gates into a one called update gate and merged cell and hidden states, as depicted in Figure~\ref{fig:gru_neuron}.

The aforementioned related works targeting short-term trajectory prediction for full-immersive VR with redirected walking generally utilize LSTMs, while here we also consider GRU networks due to their faster learning and execution times.
Moreover, in the literature the near future trajectory predictions are predominantly based on previous locations and historical movements patters of the users, taking their inspiration from works targeting the prediction of walking trajectories assuming natural human walk.
However, in full-immersive VR setups with redirected walking the VR users' movement is not natural, as they are continuously being re-steered by the redirected walking for collision avoidance.
To do so, redirected walking will deliver VR content to the users in such a way that the users are drawn toward physical locations where there is no collision hazard, while has two important implications. 
First, as the movement trajectories of VR users do not resemble a natural walk, the approaches developed for predicting near-future trajectories based on historical locations assuming naturally walking users are unoptimized for full-immersive VR with redirected walking.
Second, the redirections that are envisioned to occur in the virtual world in the near future to avoid collisions are the defining feature of the mobility of VR users in the physical setups.

\begin{figure}[!t]
\centering
\includegraphics[width=0.98\linewidth]{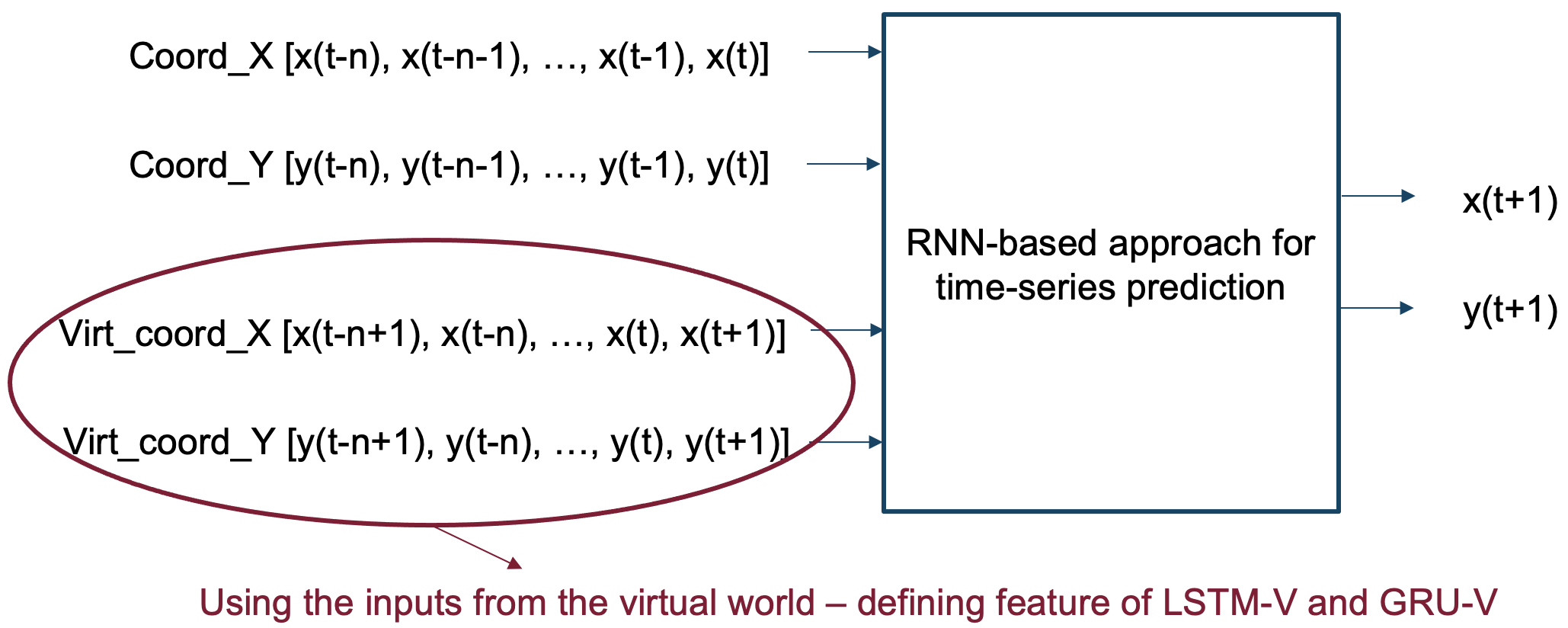}
\caption{Input features of the considered RNN approaches}
\label{fig:virtual_inputs}
\vspace{-5mm}
\end{figure}

Our intuition is that the redirected walking-related inputs from the virtual worlds are a valuable source of information for optimizing the near-future movement trajectory prediction for full-immersive VR users. 
To model such information, we define a notion of a virtual location of the VR user, i.e., its location in the virtual space.
We then utilize the stream of historical virtual locations of the VR users as an input feature to the LSTM and GRU networks, in addition to the stream of their historical physical locations. 
Based on the type of utilized input information, we distinguish the ``baseline'' (i.e., LSTM/GRU-B) and ``virtual'' (i.e., LSTM/GRU-V) versions of the utilized RNN approaches. 
The former is inspired by the existing literature, while the latter additionally accounts for the specificities of the VR users' mobility by utilizing the users' virtual coordinates.
Also note that, given that the decisions stemming from redirected walking, and consequently also the virtual coordinates of the users, are assumed to be known one step ahead in time compared to the physical coordinates of the users, as shown in Figure~\ref{fig:virtual_inputs}.
We consider this to be a natural assumption, given that the virtual coordinates for the next time instance are in redirected walking derived using the physical coordinates from a current time instance.

%% file: results.tex
 
\vspace{-1mm}
\section{Evaluation}

\begin{figure*}[!t]
\centering
\includegraphics[width=0.85\linewidth]{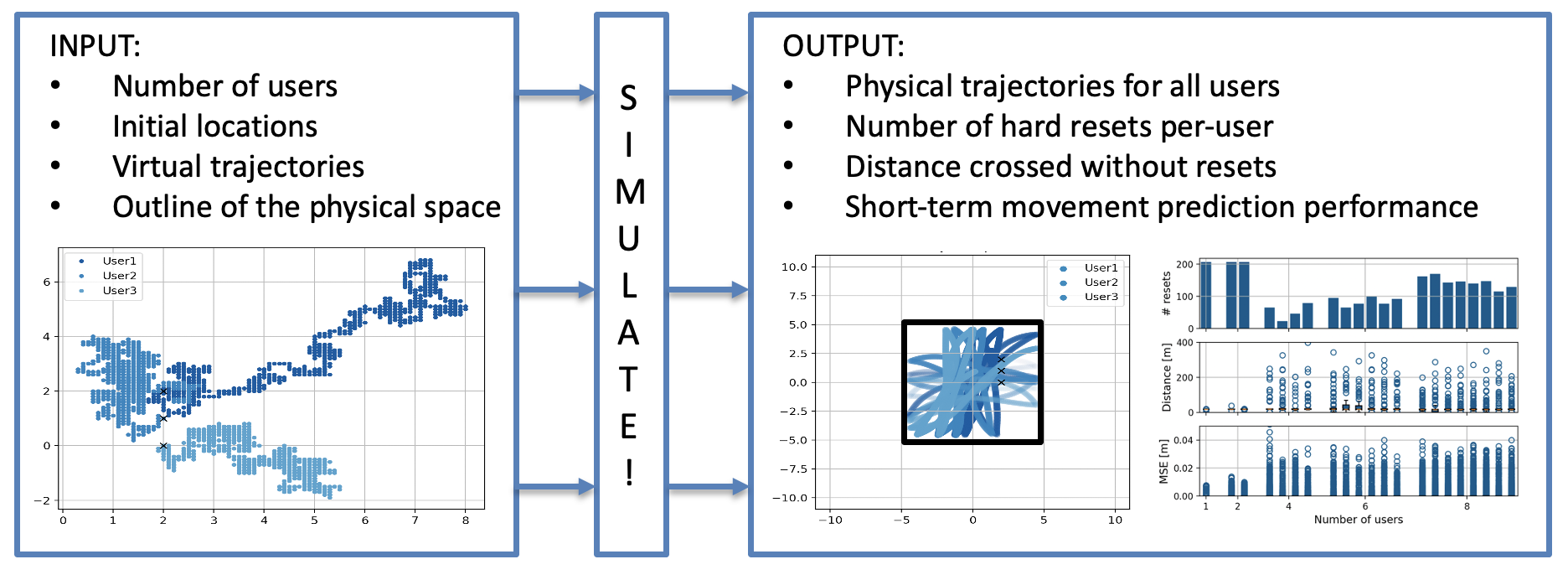}
\caption{Simulator overview~\cite{lemic2021user}}
\label{fig:simulator}
\vspace{-2.5mm}
\end{figure*} 

\subsection{Evaluation Setup}

To simulate the users' mobility in multiuser VR setups with redirected walking, we utilize the simulator from~\cite{lemic2021user}.
The core idea of the simulator is to provide a mapping between design decisions about the users' mobility in the virtual words, and their effects on the physical mobility in constrained VR environments. 
Specifically, given a set of VR users with their virtual movement trajectories, the outline of the physical environment, and a redirected walking algorithm for avoiding physical collisions, the simulator is able to derive the physical movements of the users, as indicated in Figure~\ref{fig:simulator}. 
Based on the derived physical movements, the simulator is able to output a set of performance metrics characterizing the number of perceivable resets and the distances between such resets for each user. 
We have extended the simulator to support VR users' near future movement trajectory predictions and have implemented the four considered prediction approaches (i.e., LSTM-B, LSTM-V, GRU-B, GRU-V).  
As the performance metric for characterizing the accuracy of prediction, we have defined the \ac{SE} between the predicted and true near future physical locations of the VR users.

The deployment environment is defined as a square with size of 7.5~m. 
In the training of the considered approaches, we assume two VR users coexist in the environment, fully immersed in a VR experience.
A redirected walking algorithm is then utilized to map the users' movements in the virtual world to the corresponding movements in the physical deployment environment.
We utilize the redirected walking algorithms from Bachmann~\emph{et al.}~\cite{bachmann2019multi}.
Specifically, the \ac{APF-RDW} algorithm for imperceivable redirected walking in multiuser full-immersive VR applications, and the \ac{APF-R} algorithm to re-orient the VR user towards the safest available area in case there is a need for perceivable resets to avoid collisions with other users or environmental boundaries.
Example virtual and physical mobility patterns of the users are also depicted in Figure~\ref{fig:simulator}, while the other simulator parameters are summarized in Table~\ref{tab:params}.	
The full implementation of the simulator is available as an open source tool, while the reproducibility of the results delivered in the following sections is enabled through open data access\footnote{Both available at: \url{https://bitbucket.org/filip_lemic/pm4vr/src/master/}}.

\subsection{Hyperparameter Tuning}

\begin{figure*}[!t]
\vspace{-3mm}
\centering
\subfigure[LSTM activation]{
\includegraphics[width=0.445\textwidth]{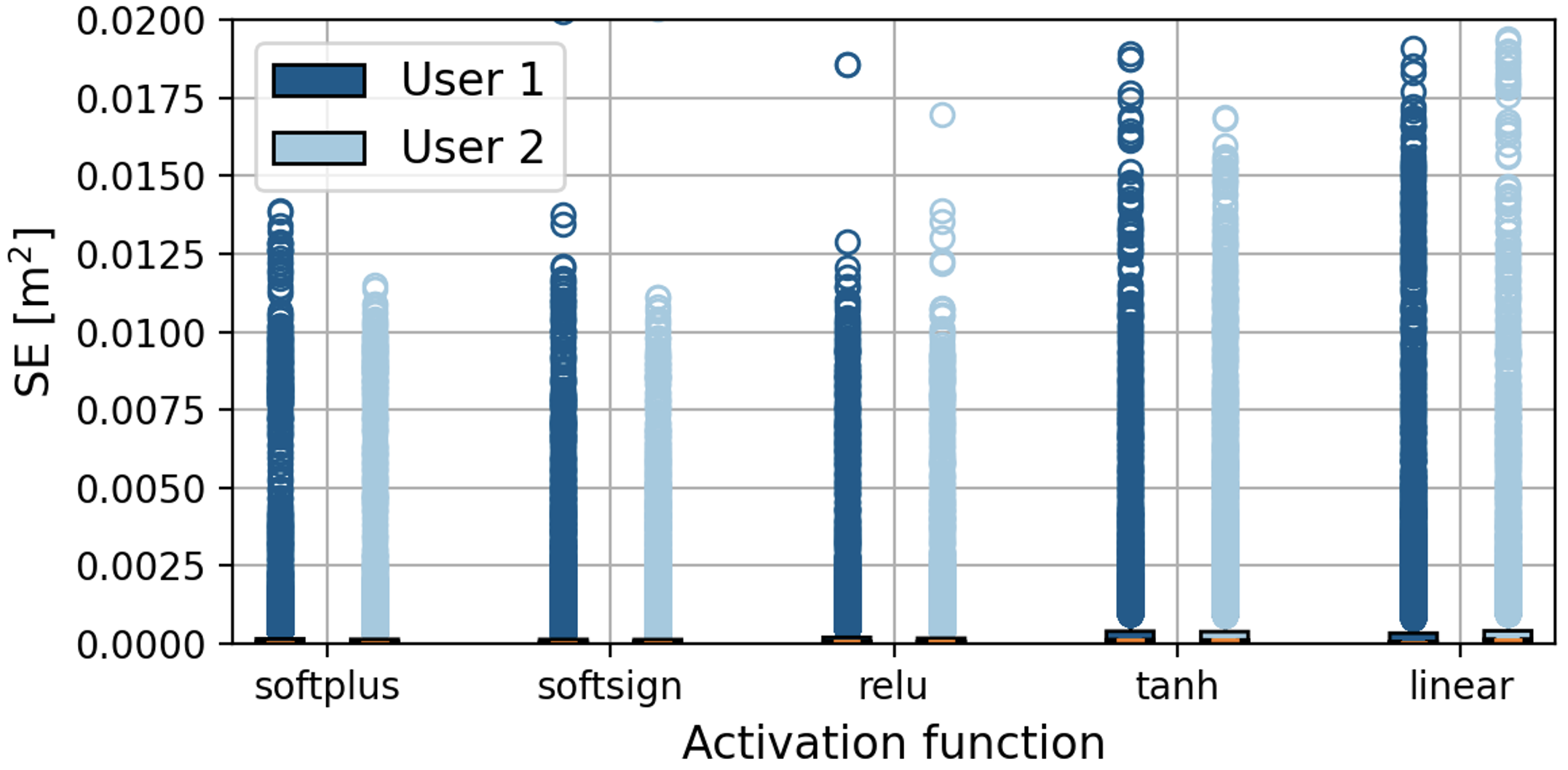}}
\subfigure[GRU activation]{
\includegraphics[width=0.445\textwidth]{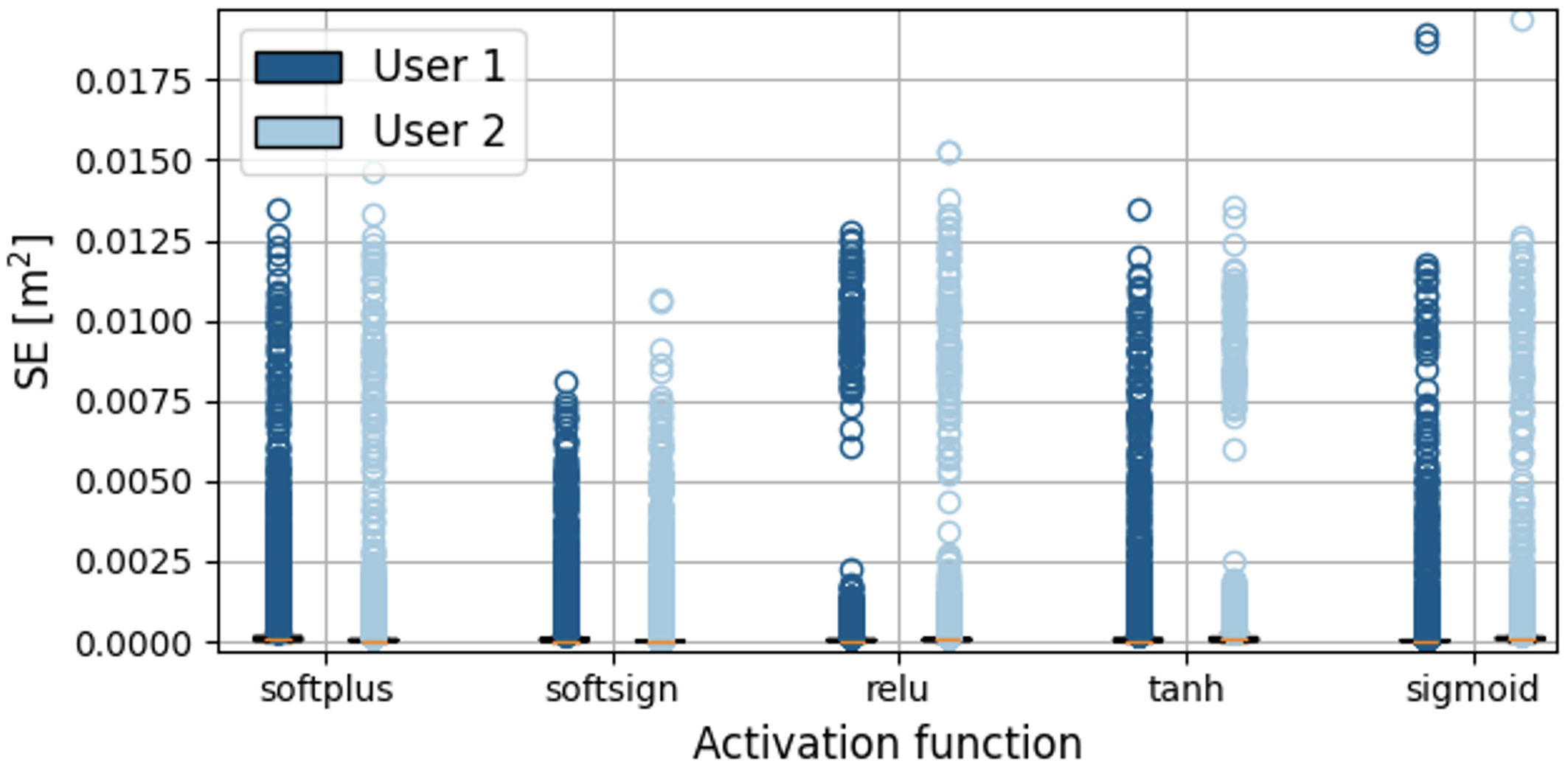}}
\subfigure[LSTM optimizer]{
\includegraphics[width=0.445\textwidth]{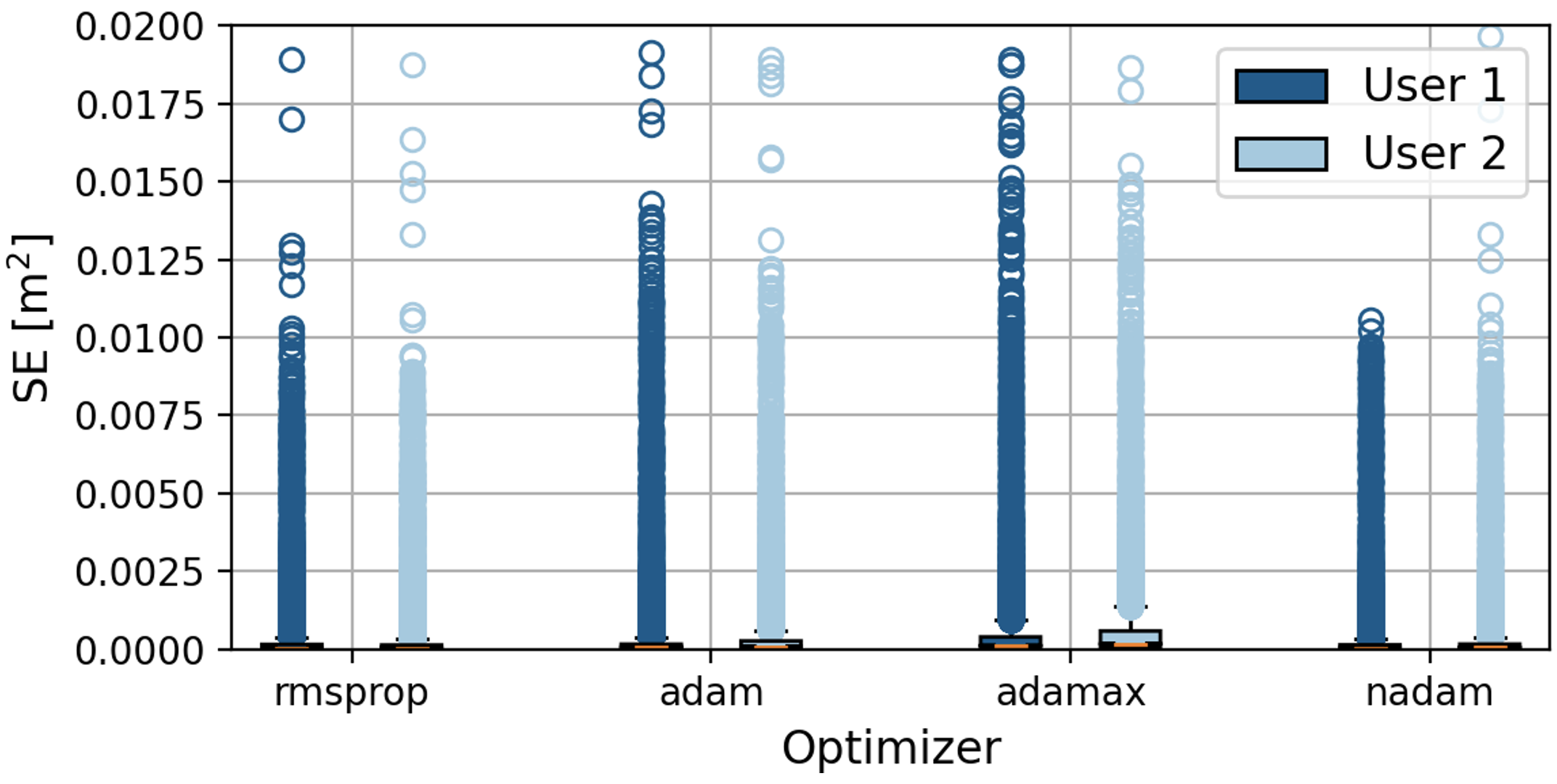}}
\subfigure[GRU optimizer]{
\includegraphics[width=0.445\textwidth]{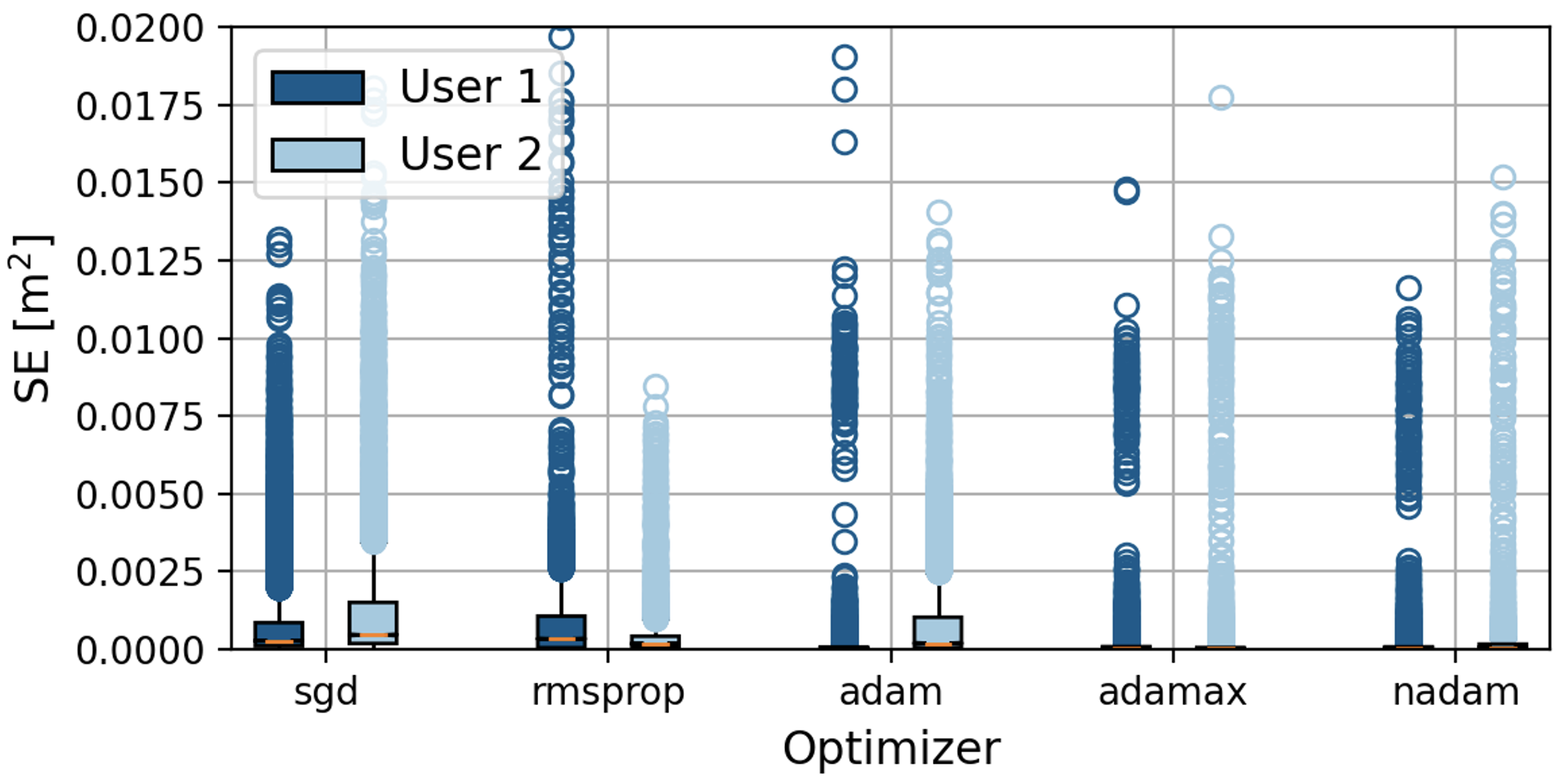}}
\subfigure[Number of LSTM neurons]{
\includegraphics[width=0.445\textwidth]{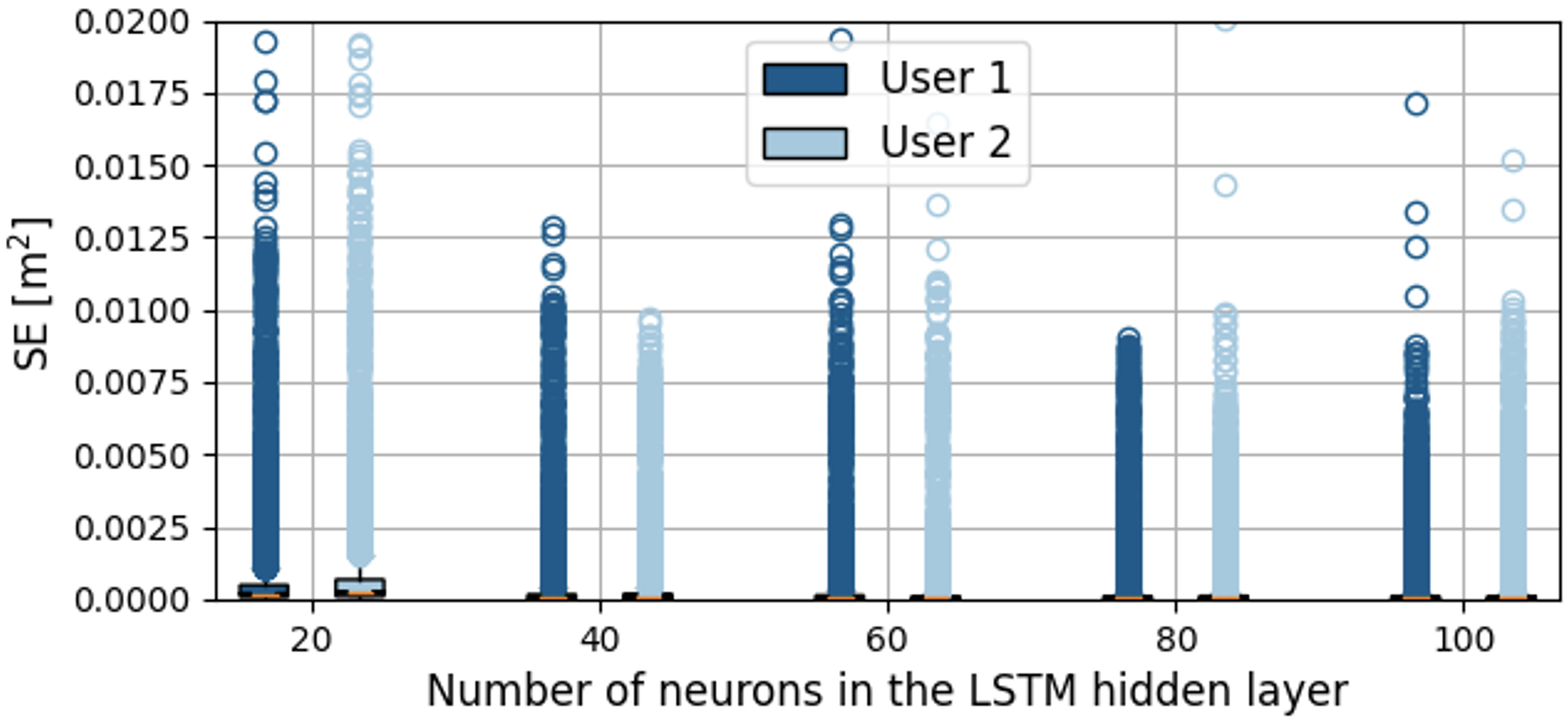}}
\subfigure[Number of GRU neurons]{
\includegraphics[width=0.445\textwidth]{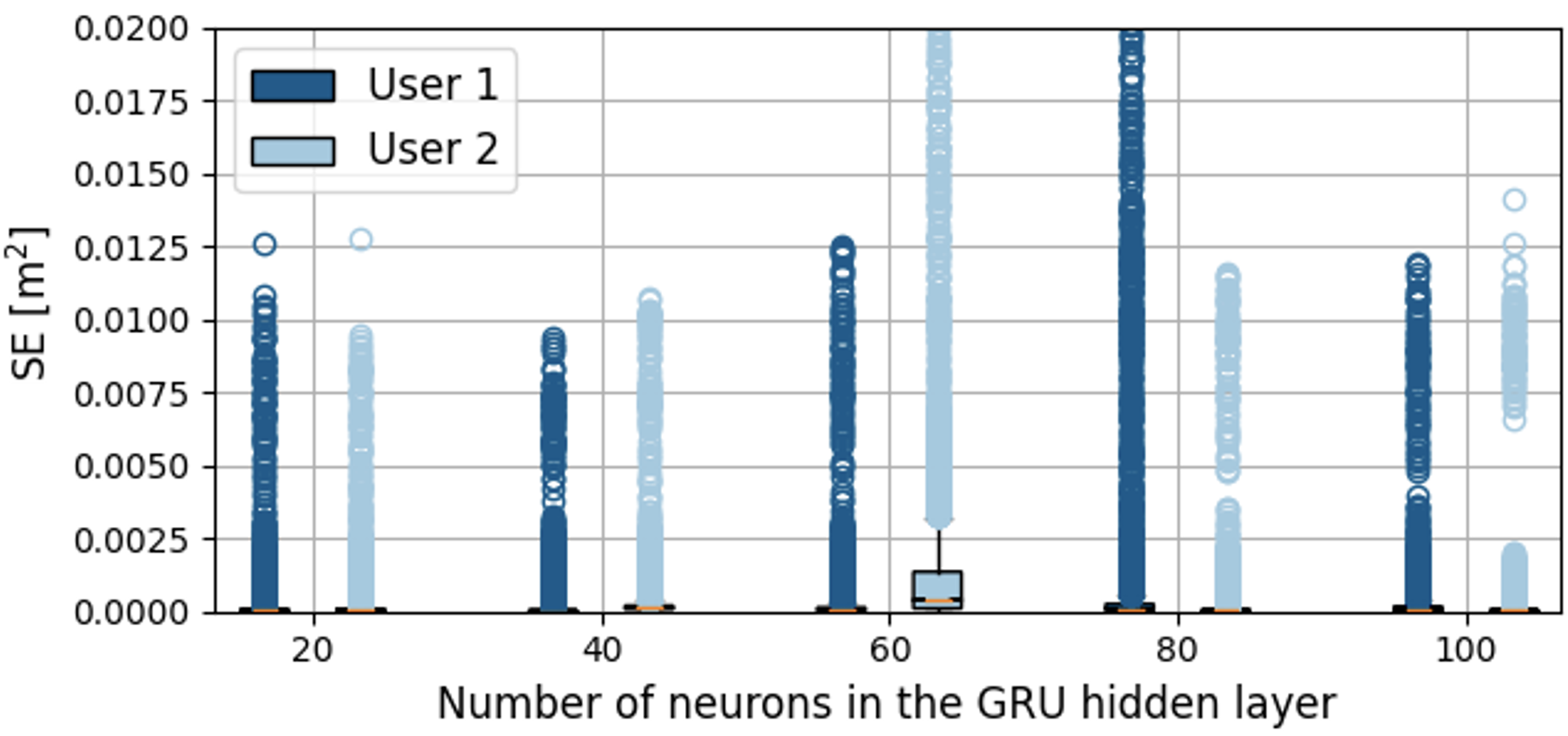}}
\vspace{-1.5mm}
\caption{Selected hyperparameter tuning results for the considered RNNs}
\label{fig:hyperparameters}
\vspace{-5mm}
\end{figure*} 

\begin{table}
    \centering
    \vspace{-1mm}
    \caption{Simulation parameters}
    \vspace{-2mm}
    \label{tab:params}
    \begin{tabular}{ l p{2.5cm}}
    \textbf{Parameter} & \hfil \textbf{Value} \\ \hline
    Number of points per trajectory & \hfil 10 [points/sec] \\
    Duration of simulation & \hfil 3600 [sec] \\ \hline
    \multicolumn{1}{c}{\hfill \textbf{Redirected walking}} \\
    Exponential falloff due to other users & \hfil 1.4 \\
    Arc radius or redirected walking & \hfil 7.5 [m] \\
    Maximum rotational rate & \hfil 15 [\textdegree] \\ 
    Velocity threshold & \hfil 0.1 [m/s] \\ \hline
    \end{tabular}
    \vspace{-1.5mm}
\end{table}

\begin{table}[!t]
\caption{Hyperparameter values used in RNNs' training}
\begin{center}
\vspace{-2mm}
\label{tab:hyperparameters}
\begin{tabular}{l c c c c c} 
\textbf{Parameter} & \textbf{Initial} & \textbf{LSTM-B} & \textbf{LSTM-V} & \textbf{GRU-B} & \textbf{GRU-V} \\
\hline
Activation & \emph{softplus} & \emph{relu} & \emph{relu} & \emph{softsign} & \emph{softmax} \\
Optimizer & \emph{sgd} & \emph{nadam} & \emph{adam} & \emph{nadam} & \emph{nadam} \\
\# neurons & 20 & 80 & 80 & 40 & 80 \\ 
\# batch/epochs & 20/10 & 20/40 & 60/50 & 80/30 & 80/30 \\
\hline
\end{tabular}
\end{center}
\vspace{-6mm}
\end{table}

In this step of the evaluation, our goal is to determine which hyperparameters have a considerable effect on the performance of the considered RNN approaches. 
We do that with the aim of reducing the hyperparameters’ optimization space, i.e., the hyperparameter tuning would be needed only for influential hyperparameters. 
We define influential hyperparameters as the ones whose change affects the average prediction error by more than 10\%, specifically the activation function, optimizer, number of neurons in the hidden LSTM/GRU layer, and number of batches and epochs. 
For the influential hyperparameters, our consequent goal is to determine the hyperparameterizations that yield optimal performance for a given approach for the near-future movement trajectory prediction in full-immersive multiuser VR setups with redirected walking. 
In our optimization approach, we tune the influential hyperparameters individually, while keeping the ones not already tuned at their initial values (cf., Table~\ref{tab:hyperparameters}).
Hence, we are able to derive a set of locally optimal hyperparameters for each of the considered approaches, as shown in the table. 
We follow this procedure, in contrast to performing an extensive multihyperparameter search (which would result in a globally optimal hyperparameters) because of an extremely high computational complexity and time overhead of the alternative.
A selected set of hyperparameter tuning results is depicted in Figure~\ref{fig:hyperparameters} for the LSTM-B and GRU-B approaches, while the other hyperparameters and considered approaches (i.e., LSTM-V, GRU-V) are left out for brevity.
Note that in the figure we depict only the hyperparameter values with ``reasonable'' prediction accuracies, while the ones with comparatively larger prediction errors are left out for clarity.  
The hyperparameter value that maximizes the prediction accuracy is then considered as the optimal one.
For example, the optimal activation functions of the LSTM-B and GRU-B approaches are respectively \emph{relu} and \emph{softsign}, as they yielded the smallest SE of the prediction, as shown in the figure.
The resulting set of hyperparameters for the four considered approaches is given in Table~\ref{tab:hyperparameters}.

\subsection{Evaluation Results}

\begin{figure}[!t]
\centering
\includegraphics[width=0.98\linewidth]{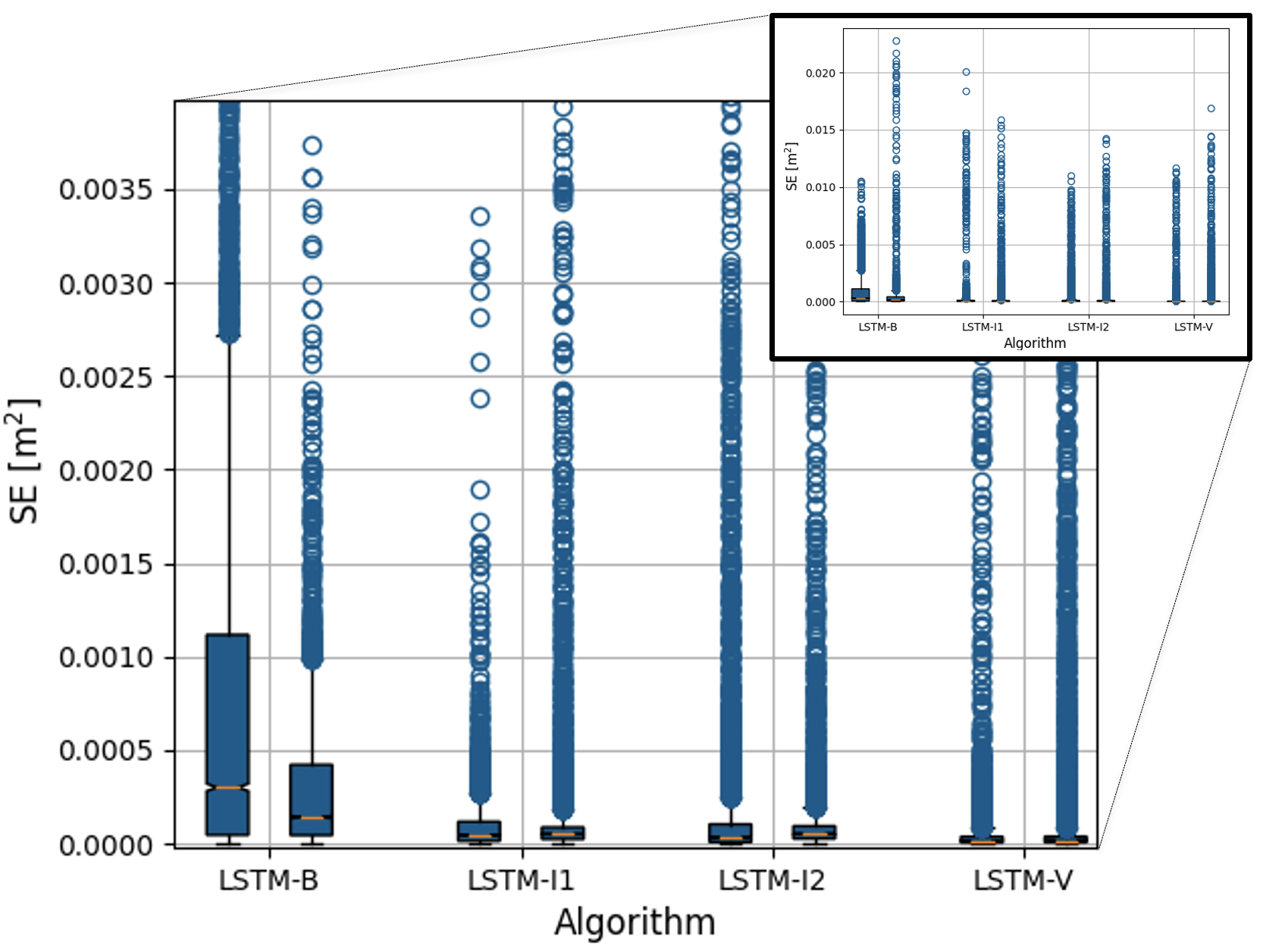}
\vspace{-2mm}
\caption{SEs achieved by different versions of the LSTM approach}
\label{fig:mse_lstm}
\vspace{-3mm}
\end{figure}

\begin{figure}[!t]
\centering
\includegraphics[width=0.98\linewidth]{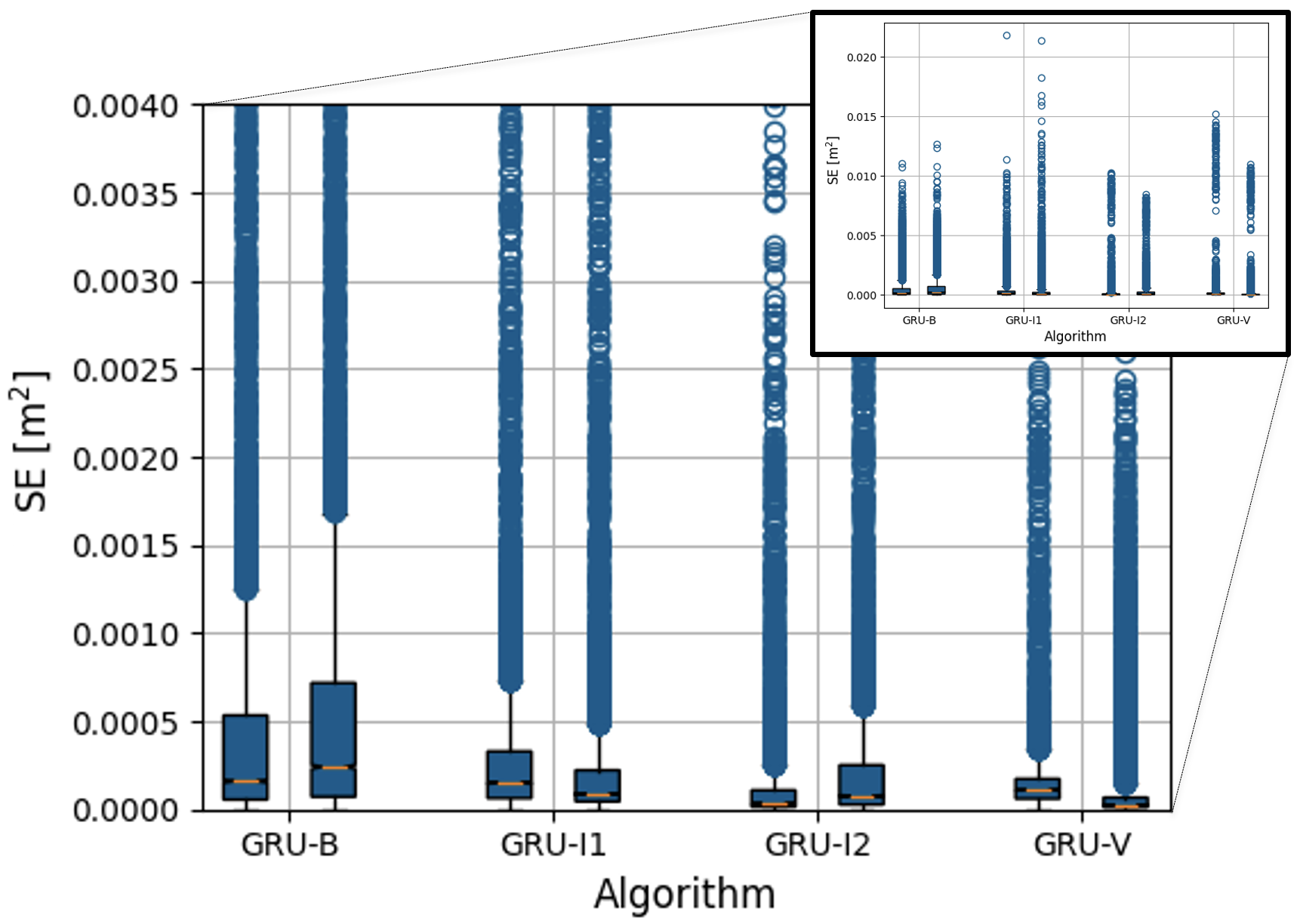}
\vspace{-2mm}
\caption{SEs achieved by different versions of the GRU approach}
\label{fig:mse_gru}
\end{figure}

We present the performance results achieved by the four considered prediction approaches, with their hyperparameters optimized following the above-discussed procedure. 
Figures~\ref{fig:mse_lstm} and~\ref{fig:mse_gru} depict the \acp{SE} observed by different versions of the LSTM and GRU-based prediction approaches, respectively.
For each approach, we depict the baseline performance (i.e., GRU/LSTM-B), as well as the performance of their virtual counterparts (i.e., LSTM/GRU-V), both with locally-optimal hyperparameters. 
Moreover, we also depict two intermediary approaches.
In LSTM/GRU-I1, the hyperparameters of a predictor have not been changed in comparison to the baseline approach to isolate solely the effect of utilizing virtual trajectory as a feature for prediction.
In LSTM/GRU-I2, only the number of neurons in the hidden layer and the number of epochs have been doubled in comparison to the baseline approach. This has been done with an intuition of better handling the complexity of utilizing virtual coordinates as an additional input feature compared to LSTM/GRU-I1.

As visible in the figure, for both LSTM- and GRU-based approaches, the versions utilizing the virtual coordinates generally outperform the baselines based on utilizing solely historical knowledge about the physical movement trajectories of the VR users.
Additionally, there are also benefits of hyperparameter tuning when utilizing virtual coordinates as an additional input feature. 
For example, the average per-user \ac{SE} of the prediction is deceased from roughly 0.0003~m$^2$ in LSTM-B to less than 0.0001~m$^2$ in LSTM-V, representing a threefold increase in the prediction accuracy.
Moreover, a roughly fivefold improvement is observed for the GRU-based prediction when utilizing virtual coordinates of the VR users in comparison to the more traditional baseline.
We argue that these results demonstrate that certain contextual information from the virtual world, in this case the virtual coordinates of the VR users, can be utilized for improving the performance of the RNN-based near-term movement trajectory prediction approaches in the physical VR setups.
We believe that this conclusion can be generalized and the fact that we have demonstrated the benefits for two types of prediction approaches makes an argument for supporting this intuition. 

\begin{figure}[!t]
\centering
\includegraphics[width=0.9\linewidth]{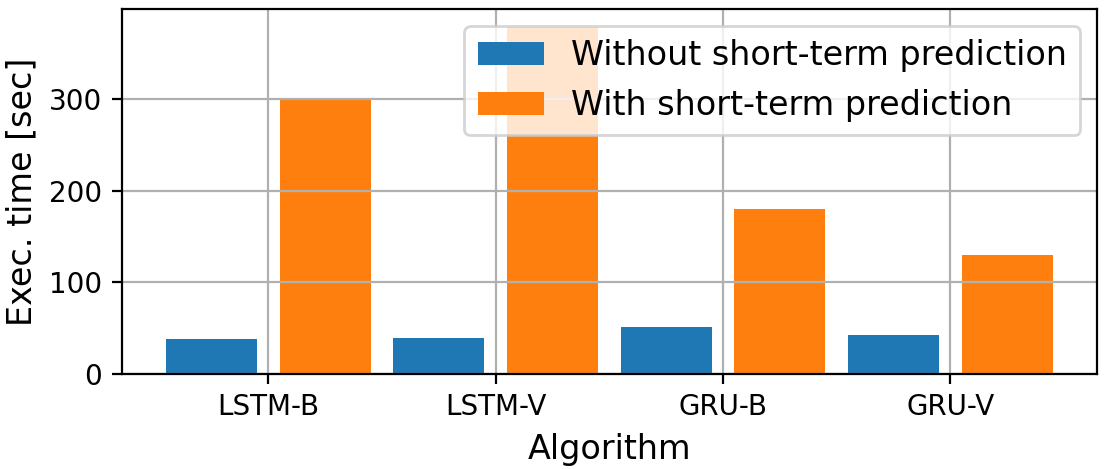}
\vspace{-1mm}
\caption{Training time of different approaches}
\label{fig:times}
\vspace{-3mm}
\end{figure}  

The training duration of the four considered approaches is depicted in Figure~\ref{fig:times}. 
Note that the execution times have been captured using the simulator running on a contemporary MacBook Air laptop and as such their absolute values are indicative solely to that platform.
Nonetheless, several interesting trends can be observed.
The figure depicts the execution times of different approaches in cases when the short-term trajectory prediction is enabled and vice-versa.
Intuitively, where the trajectory prediction is not enabled, the simulation times of both ``baseline'' and ``virtual''approaches are comparable, as shown in the figure.
This is an expected behaviour as the prediction approaches are in this case neither trained nor involved in the prediction, but it nonetheless makes an argument about the correctness of the implementation.

In case the short-term trajectory prediction is enabled, one can first observe that the GRU-based approaches generally features lower simulation times than the LSTM-based alternatives.
This is a well-known behaviour demonstrating one of the primarily benefits of GRU over LSTMs, namely their faster execution and training. 
Finally and perhaps counterintuitively, the ``virtual'' version of the LSTM-based approach features higher simulation time than its 'baseline' counterpart, while that is not the case for GRU-flavored prediction methods. 
We argue this discrepancy comes from the fact that an increase in the number of input features of the ``virtual'' version of both approaches is not the primary factor in their execution times, despite the fact that it is certainly to an extent contributive to the increase in the execution times.
On the contrary, we argue that the execution times are to a greater extent affected by the hyperparameterization of the approaches, with the most significant contributions stemming from the selected number of batches, epochs, and neurons.
In case of LSTM, the comparatively higher number of sample per batch in LSTM-V allows for greater parallelization of the training process, which arguably has a larger effect on the execution time than the combined effect of a slight increase in the number of training epochs and the increase in the input feature dimensionality compared to LSTM-B, hence the execution time of LSTM-V is lower than for LSTM-B.
In case of GRU, an increase in the number of neurons in the hidden GRU layer and in the dimensionality of the input features jointly affect the increase in the execution time of GRU-V compared to GRU-B.
In conclusion, we argue the increase in the dimensionality of the input features when considering virtual coordinates has only a slight negative effect on the training and execution times of the considered approaches, while a more substantial affect can be expected from hyperparameter tuning.

\begin{figure*}[!t]
\vspace{-1mm}
\centering
\subfigure[LSTM-V]{
\includegraphics[width=0.44\linewidth]{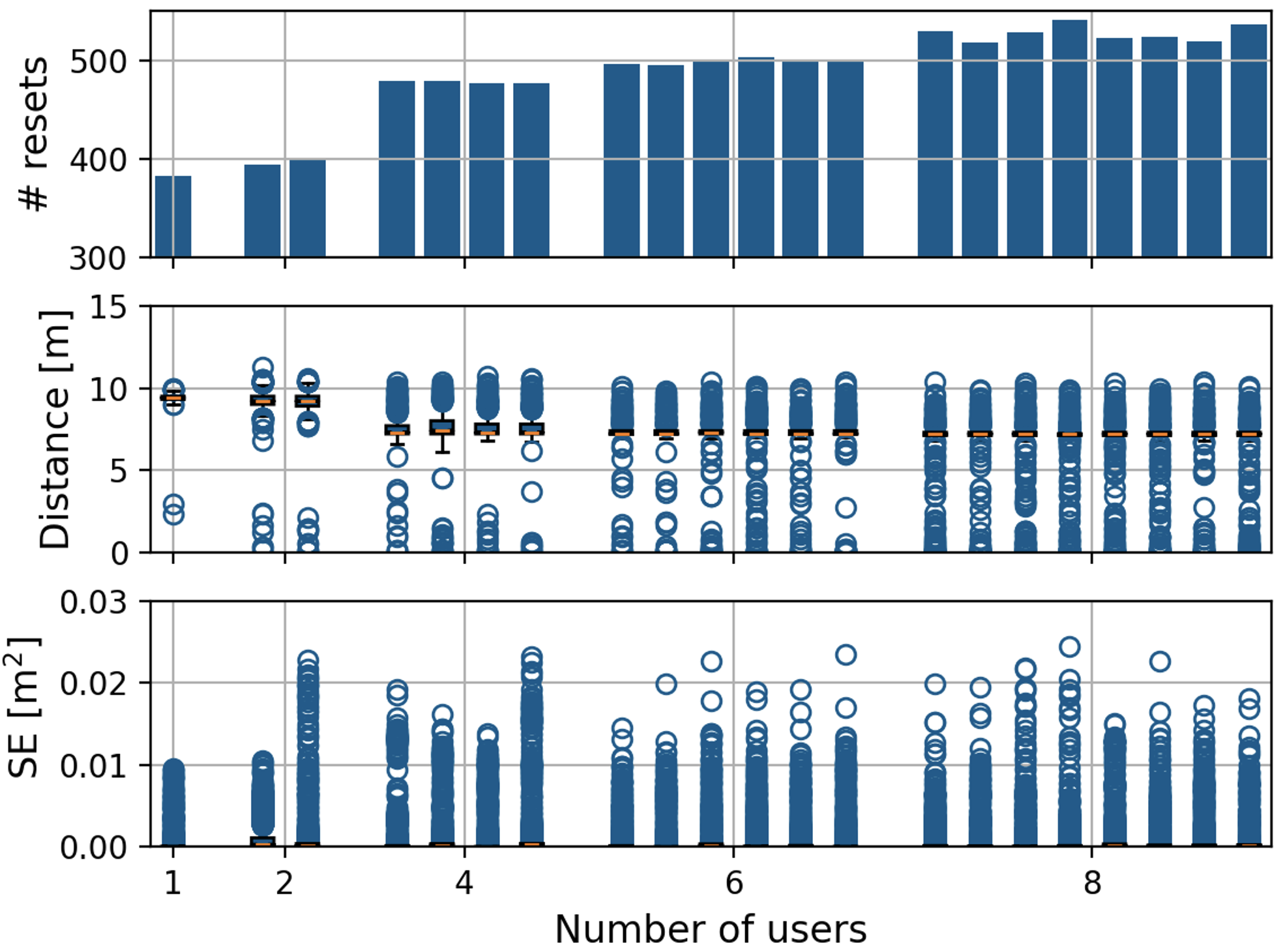}}
\subfigure[GRU-V]{
\includegraphics[width=0.44\linewidth]{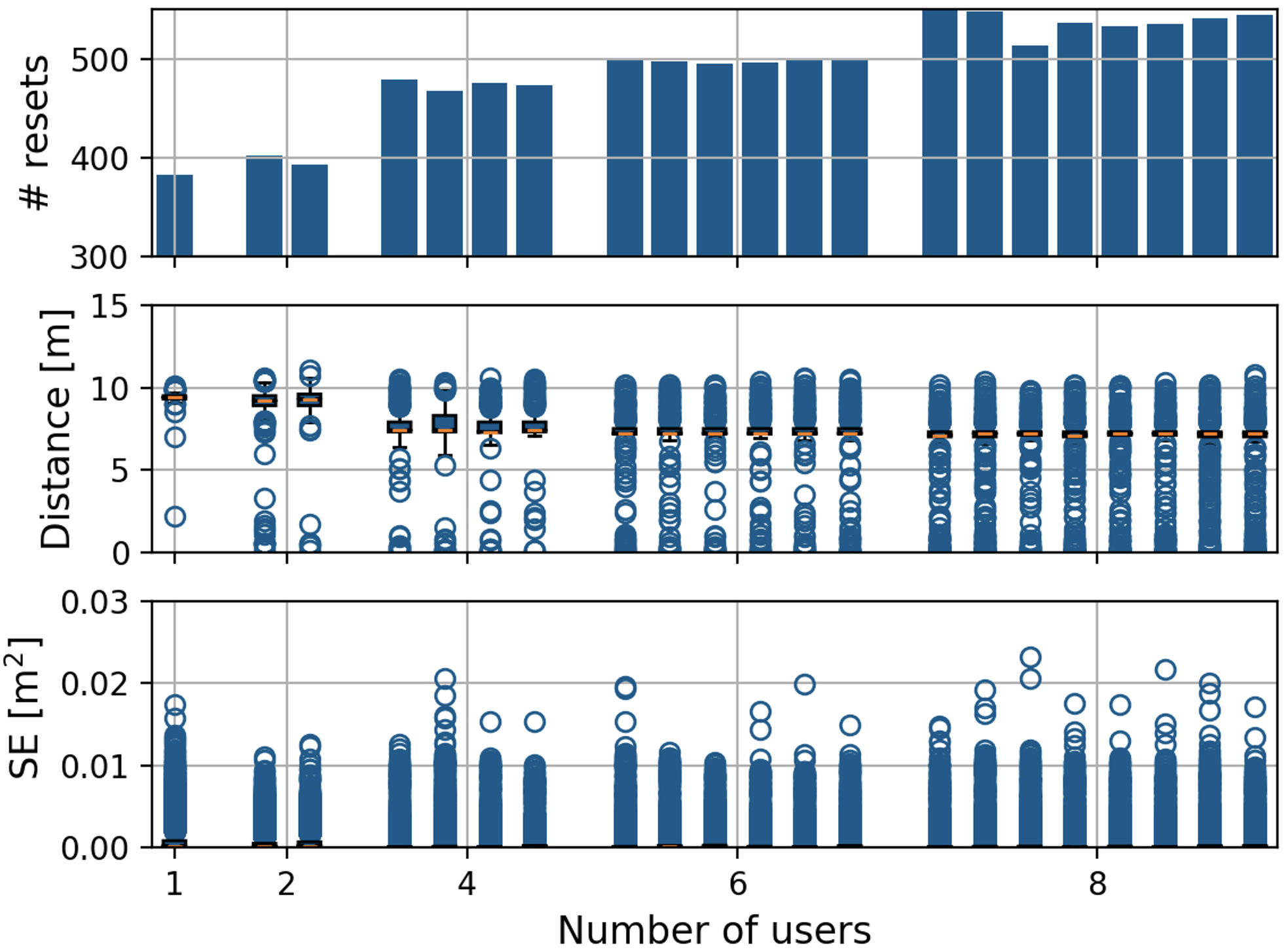}}
\vspace{-2mm}
\caption{Multiuser performance results for LSTM-V and GRU-V}
\label{fig:multiuser}
\vspace{-5.5mm}
\end{figure*} 

Finally, the extended simulator enables straightforward evaluation of the performance of redirected walking and short-term trajectory prediction for a varying number of coexisting VR users.
We utilize this feature to evaluate if the considered approaches can, once optimally trained for a two-user system, be used for such prediction for a varying number of coexisting users.
Figure~\ref{fig:multiuser} depicts the output of the simulator (i.e., a set of performance metrics) for the two optimally-parameterized prediction approaches for different numbers of coexisting VR users.
The graphs in the first row depict the number of perceivable resets experienced by each VR user for a varying number of the VR users.
Similarly, the graphs in the second row depict the distances the users are able to traverse between two consecutive perceivable resets.
As visible from the figure, the numbers of perceivable resets and the distances between them are highly comparable for the two utilized approaches.
This is an expected behaviour as the same redirected walking approach, which has an affect on the number of perceivable rests and their respective distances, has been utilized for both shot-term trajectory prediction approaches.
Moreover and as visible in the figure, the number of such resets increases (and, therefore, the corresponding distances between consecutive resets decrease) with an increase in the number of users, which is again an expected behaviour as reported in~\cite{lemic2021user}. 
We depict these two metrics solely to demonstrate that there are significant differences in the performance of the prediction system as a function of a varying number of VR users.

The above suggests that datasets used for training of short-term trajectory prediction are not comparable for a varying number of users.  
Nonetheless, the prediction performance for both of the utilized approaches is highly comparable across a varying number of coexisting VR users, as indicated int he bottom graphs of the figure depicting the \acp{SE} of predictions.
This suggests that both systems could, without significant prediction performance degradation, be used in scenarios in which the users dynamically immerse in or terminate shared VR experiences.
In addition, one can observe that the GRU-V approach generally outperforms LSTM-V, with average reduction in the observed SEs of around 50\%, suggesting their utility for the movement trajectory prediction in full-immersive multiuser VR setups with redirected walking.

%% file: conclusion.tex
\section{Conclusion}
\label{sec:conclusion}

We have shown that \acf{LSTM} and \acf{GRU}-based \acfp{RNN} are promising candidates for the near-future movement trajectory prediction in multiuser full-immersive \acf{VR} with redirected walking.
We have also demonstrated the benefits of utilizing virtual context such as the movement trajectory in the virtual world as an input feature for such prediction, as well as shown that that the prediction system, once trained on assuming on a static number of users, can maintain the prediction accuracy once the number of users in the system changes, without the need for re-training.
The main direction of our ongoing and future work is and will be focused on experimentally confirming our insights with realistic traces.